\documentclass[preprint,12pt]{elsarticle}
\usepackage{multirow}
\usepackage{array}
\usepackage{graphicx}

\usepackage{amssymb}

\journal{Mathematics and Computers in Simulation}

\begin{document}

\begin{frontmatter}


  \author{Igor~Zacharov}
  \author{Rinat~Arslanov}
  \author{Maksim~Gunin}
  \author{Daniil~Stefonishin}
  \author{Sergey~Pavlov}
  \author{Oleg~Panarin}
  \author{Anton~Maliutin}
  \author{Sergey~Rykovanov}
  \author{Maxim~Fedorov}

  \address{Skolkovo Institute for~Computational and Data-Intensive Science and Engineering~(CDISE) 143026 Moscow, Russia}

\title{``Zhores''~\textemdash Petaflops supercomputer for~data-driven modeling, machine learning and~artificial~intelligence installed in Skolkovo Institute of Science and Technology}


\begin{abstract}

The Petaflops supercomputer ``Zhores'' recently launched in~the~``Center for Computational and Data-Intensive Science and Engineering''~(CDISE) of Skolkovo Institute of Science and Technology (Skoltech) opens up new exciting opportunities for~scientific discoveries in~the~institute especially in the areas of data-driven modeling, machine learning and artificial intelligence. This supercomputer utilizes the latest generation of~Intel and NVidia processors to~provide resources for~the~most compute intensive tasks of the Skoltech scientists working in digital pharma, predictive analytics, photonics, material science, image processing, plasma physics and many more. Currently it places 6$^{th}$ in~the~Russian and CIS TOP-50~(2018) supercomputer list. In this article we summarize the cluster properties and discuss the measured performance and usage modes of~this~scientific instrument in~Skoltech. 
\end{abstract}

\end{frontmatter}


\section{Introduction}
\label{Introduction}

Skoltech CDISE Petaflops supercomputer ``Zhores" named after the Nobel Laureate Zhores Alferov, is intended for cutting-edge multidisciplinary research in data-driven simulations and modeling, machine learning, Big Data and artificial intelligence (AI). It enables research in such important fields as Bio-medicine~\cite{zhang2018multi, yokota2018semi}, Computer Vision~\cite{kononenko2015learning, kononenko2018semi, burkov2018deep, ulyanov2018deep, yurchenko2017parsing, babenko2014neural}, Remote Sensing and Data Processing~\cite{mirvakhabova2018field, novikov2018satellite, cao2018robust}, Oil/Gas~\cite{muravleva2018application, klyuchnikov2018data}, Internet of Things~\cite{somov2018bacteria, somov2018pervasive}, High Performance Computing (HPC)~\cite{matveev2018parallel, cichocki2018tensor}, Quantum Computing~\cite{biamonte2017quantum, baez2018quantum}, Agro-informatics~\cite{mirvakhabova2018field}, Chemical-informatics~\cite{0953-8984-30-32-32LT03,novikov2018improving, docampo2016molecular, coles2017nanostructure, gomez2017molecular} and many more. Its architecture reflects the modern trend of convergence of ``traditional'' HPC, Big Data and AI. Moreover, heterogeneous demands of Skoltech projects on computing possibilities ranging from throughput computing to capability computing and the need to apply modern concepts of workflow acceleration and in-situ data analysis impose corresponding solutions on the architecture. The design of the cluster is based on the latest generation of CPUs, GPUs, network and storage technologies, current as of 2017--2019. This paper describes the implementation of this machine and gives details of the initial benchmarks that validate its architectural concepts.

The article is organized as follows. In section~\ref{sec:install} the details of installation are discussed with subsections dedicated to the basic technologies. Section~\ref{sec:apps} describes several applications ran on the ``Zhores'' cluster and their scaling. The usage of the machine in the~``Neurohackaton'' held in November 2018 in Skoltech is described in section~\ref{sec:Neurohackathon}. Finally, section~\ref{sec:Conclusions} provides conclusions.

\section{Installation}
\label{sec:install}
\begin{figure}[h!]
    \centering
    \includegraphics[width=\columnwidth]{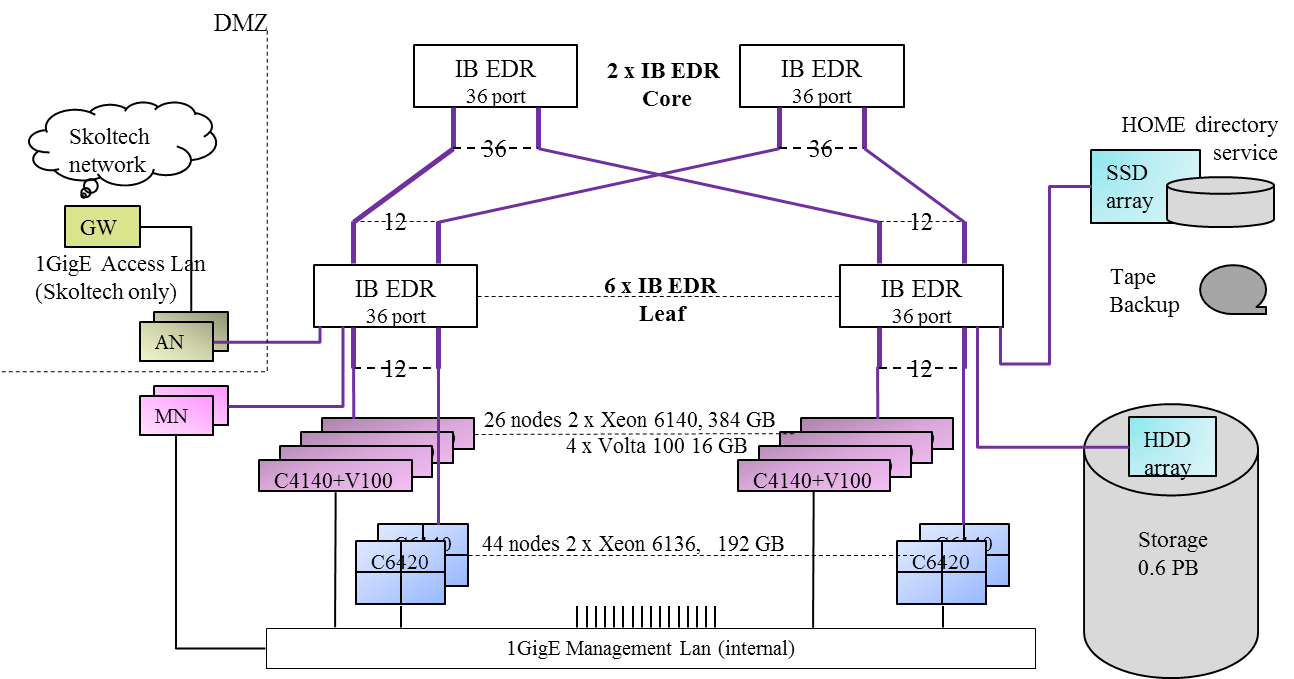}
    \caption{Principle connection scheme. The~an and~mn nodes are marked explicitly; the cn, gn and other nodes are lumped together.}
    \label{fig:principle_scheme}
\end{figure}

''Zhores'' is constructed from the DELL PowerEdge C6400 and C4140 servers with Intel\textregistered{} Xeon\textregistered{} CPUs and Nvidia Volta GPUs connected by Mellanox EDR Infiniband (IB) SB7800/7890 switches. We decided to allocate 20~TB of the fastest storage system (based on NVMe over IB technology) for small users' files and software (home directories), and 0.6 PB GPFS file system for~bulk data storage. The principal scheme with the majority of components is illustrated in fig.~\ref{fig:principle_scheme}.
The exact composition with the characteristics of the components is found in table~\ref{tab:ZHORES_components}. The names of the nodes are given according to their intended role:
\begin{itemize}
    \item 
    cn~--- compute nodes to handle CPU workload
    \item
    gn~--- compute nodes to handle GPU workload
    \item
    hd~--- hadoop nodes with set of disks for the classical Hadoop workload
    \item
    an~--- access nodes for cluster login, submit jobs and transfer users' data
    \item
    anlab~--- special nodes for user experiments
    \item
    vn~--- visualization nodes
    \item
    mn~--- main nodes for cluster management and monitoring
\end{itemize}

All users land on one of the access nodes (an)~after login and can use them for interactive work, data transfer and for job submission (dispatching tasks to compute nodes). Security requirements place the access nodes in the demilitarized zone. The queue structure is implemented using the~SLURM workload manager and discussed in section~\ref{sec:Queues}. Both, shell scripts and Docker~\cite{Docker} images are accepted as valid work item by the queuing system.
We have made a principal decision to use the latest CentOS version~7.5 which was officially available at the time of installation. The user environment is provided with the Environment Modules software system~\cite{Software_modules}.
Several compilers (Intel and GNU) are available as well as different versions of pre-compiled utilities and applications.

The cluster is managed with the fault tolerant installation of the Luna management tool~\cite{trinityX}. The two management nodes are mirrors of each other and provide the means of provisioning and administration of the cluster, provide the~NFS export of user /home directories and all cluster configuration data. This is described in section~\ref{sec:Luna}.

\subsection{Servers' Processor Characteristics}
The servers have the latest generation of the Intel Xeon processors and Nvidia Volta GPUs. The basic characteristics of each type of the servers are captured in table~\ref{tab:ZHORES_components}. We have measured the salient features of these devices.

\begin{table}[ht!]
    \centering
    \caption{Details of named~``Zhores'' cluster nodes}
    \label{tab:ZHORES_components}
    \begin{tabular}{|l|l|>{\centering}p{14mm}|>{\centering}p{9mm}|>{\centering}p{14mm}|>{\centering}p{12mm}|r|r|r|}
        \hline
        Name & CPU & sockets $\times$ cores & $F$ [GHz] & Memory [GB] & Storage [TB]  &  [TF/s]   & \#  &[TF/s] 
        \tabularnewline
        \hline
        cn     & 6136 & 2 x 12   & 3.0  & 192  & 0.48  &  2.3 & 44 & 101.4
        \tabularnewline
        \hline 
        \multirow{2}{*}{gn} & 6140 & 2 x 18   & 2.3  & 384  & 0.48  & 2.6 & \multirow{2}{*}{26} & 68.9
        \tabularnewline
        & V100 & \hspace{-1mm}4 x 5120 & 1.52 & 4 x 16 &    & 31.2  &   & 811.2
        \tabularnewline
        \hline
        hd    & 6136 & 2 x 12   & 3.0  & 192  & 9.0  &  2.3  & 4 &  9.2
        \tabularnewline
        an    & 6136 & 2 x 12   & 3.0  & 256  & 4.8  &  2.3  & 2 &  4.6
        \tabularnewline
        vn    & 6134 & 2 x 8    & 3.2  & 384  &      &  1.6  & 2 &  3.2
        \tabularnewline
        anlab & 6134 & 2 x 8    & 3.2  & 192  &      &  3.3  & 4 & 13.1
        \tabularnewline
        mn    & 6134 & 2 x 8    & 3.2  &   64 &      &  3.3  & 2 &  6.6
        \tabularnewline
        \hline
        Totals  &      & 2296     &      & 21248 &       &      & 82 & 1018.2
        \tabularnewline
        \hline
    \end{tabular}
\end{table}

  Intel Xeon 6136 and 6140 ``Gold'' CPUs of~Skylake generation differ by the total number of cores in the package and the working clock frequency~($F$). Each core features two floating point AVX512 units.  This has been tested with a special benchmark to verify that the performance varies with the frequency as expected.  

\begin{figure}[h!]
    \centering
    \includegraphics[width=\columnwidth]{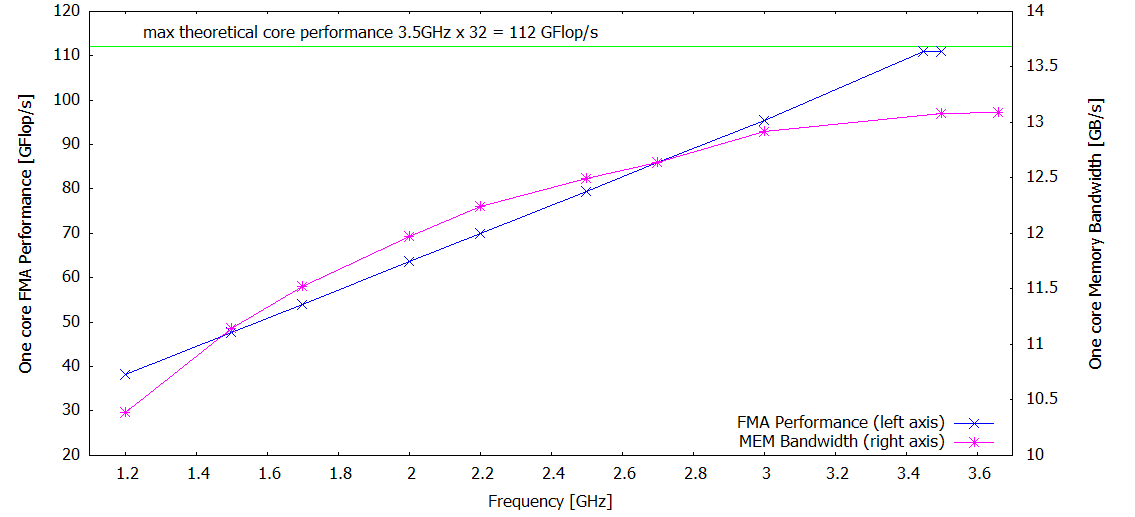}
    \caption{Floating point performance (FMA instructions) on 6136 CPU core and memory bandwidth (STREAM Triad) as a function of clock frequency. Left ordinate shows the FMA performance, the right ordinate represents the memory bandwidth.}
    \label{fig:CPU_MEM_core}
\end{figure}

The CPU performance and memory bandwidth of a single core is shown in fig.~\ref{fig:CPU_MEM_core}.
The benchmark program to test the floating point calculation performance is published elsewhere ~\cite{COLFAX}. It is an unrolled vector loop with vector width 8, precisely tuned for the AVX512 instruction set. In this loop exactly 8 double precision numbers will be computed in parallel in two execution units of each core. With two execution units and the fused multiply-add instruction (FMA) the theoretical Double Precision (DP) performance of a single physical core is $8 \times 2 \times 2 \times F$~[GHz] and for the maximum of~$F = 3.5$~GHz may reach 112~GFlop/s/core. The performance scales with the frequency to the maximum determined by processor thermal and electrical limits. 
The total FMA performance on a node when running AVX512 code on all processors in parallel is about 2.0~TFlop/s for C6140 machines (cn nodes, 24~cores) and 2.4~TFlop/s for the C4140 (gn nodes, 36~cores). Summing up all the cn and gn nodes gives the measured maximum CPU performance on the ``Zhores" cluster of 150~TFlop/s.

The latencies of the processor memory subsystem have been measured with the LMBench program~\cite{LMBench} and summarized in table~\ref{tab:CPU_Latency}.

\begin{table}[h!]
    \centering
    \caption{Memory properties from Xeon 6136/6140 processor visible from single core }
    \label{tab:CPU_Latency}
    \begin{tabular}{|l|c|c|c|c|c|l|}
    \hline
    cache     & \multirow{2}{*}{set}  &line & Latency & Bandwidh  &  size & Core \\
    level     &                     &   [Bytes]    & [ns] & [GB/s] & [KiB] & OWN\\
    \hline
     L1 Data  & 8-way & 64  & 1.1 & 58   & 32         & private\\
     L1 Instr.& 8-way         & 64  &     &      & 32         & private\\
     L2 Unif. & 16-way        & 64  & 3.8 & 37   & 1024       & private \\
     L3 Unif. & 11-way        & 64  &     &   26 & 25344      & shared \\
     TLB      &  4-way        &          &     &      & 64 entries & private \\
     \hline
     \multicolumn{3}{|l|}{Memory Xeon 6136 parts}          & 27.4 & 13.1 & 192 GB    & shared  \\
     \multicolumn{3}{|l|}{Memory Xeon 6140 parts}          & 27.4 & 13.1 & 384 GB & shared \\
     \hline
    \end{tabular}
\end{table}
The main memory performance is measured with the STREAMS program~\cite{STREAMS} and shown for the single core as a function of clock frequency in fig.~\ref{fig:CPU_MEM_core}. The theoretical performance of the memory bandwidth may be estimated with the Little Law~\cite{LittleLaw} to~14~GB/s per each channel taking into account the memory latency of 27.4~ns given in table~\ref{tab:CPU_Latency}.
The total memory bandwidth (STREAM Triad) for all cores reached 178.6~GB/s in our measurement using all 6~channels of 2666~MHz DIMMs.

The strong dependency of the FMA performance on the processor clock frequency and the weak dependency of the memory bandwidth on the clock frequency is noted to propose a scheme for the optimization of the power usage for applications with mixed instruction profiles.

%

\subsection{Nvidia V100 GPU}
\begin{figure}[h!]
    \centering
    \includegraphics[width=\columnwidth]{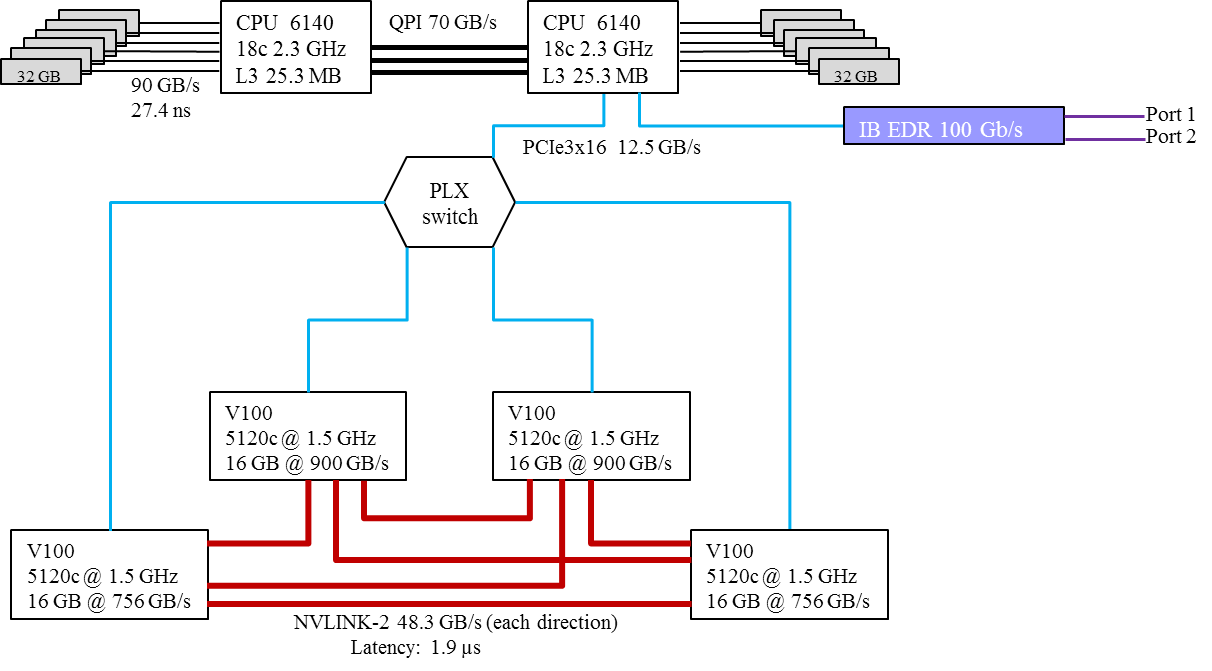}
    \caption{Principal connections between host and graphics subsystem on graphics nodes.}
    \label{fig:GPU_arch}
\end{figure}

\begin{figure}[h!]
    \centering
    \includegraphics[width=\columnwidth]{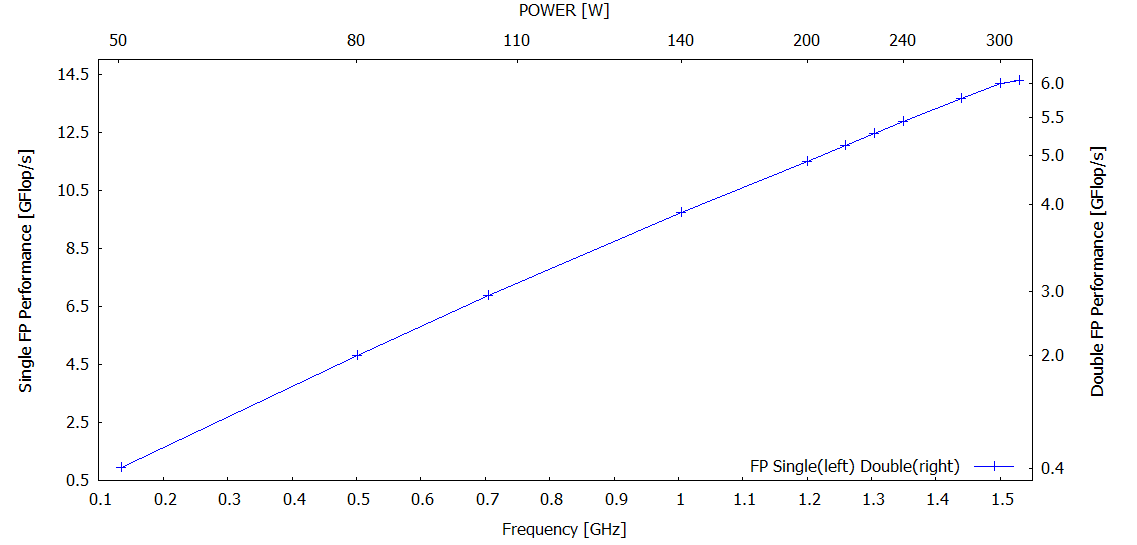}
    \caption{Nvidia V100 GPU floating point performance as a function of graphics clock rate. Electrical power draw corresponding to the set frequency is indicated on the upper axis.}
    \label{fig:GPU-POW}
\end{figure}

Significant nodes~(26) in the ``Zhores'' cluster are equipped with four Nvidia V100 GPUs each. The GPUs are connected pairwise with NVLink and individually with PCIe gen3 x16 to the CPU host. The principal scheme of the connections is shown in fig.~\ref{fig:GPU_arch}. The basic measurements to label the links in the plot have been obtained with Nvidia p2p bandwidth program from the ``Samples'' directory loaded with GPU drivers.
This setup is optimized for parallel computation scaling within the node, while the connections to the cluster network pass from the single PCIe link. 
The maximum estimated performance of a single V100 GPU is shown in fig.~\ref{fig:GPU-POW}. The graphics clock rate was set with the command ``nvidia-smi"; same command with different parameters lists the power draw of the device. The computational efficiency measured in Performance per Watt is not evenly distributed as function of frequency, the peak is 67.4~GFlop/s/W (single precision) at 1~GHz and drops to 47.7~GFlop/s/W at 1.5~GHz. 

\subsection{Mellanox IB EDR network}
\begin{figure}[h!]
    \centering
    \includegraphics[width=\columnwidth]{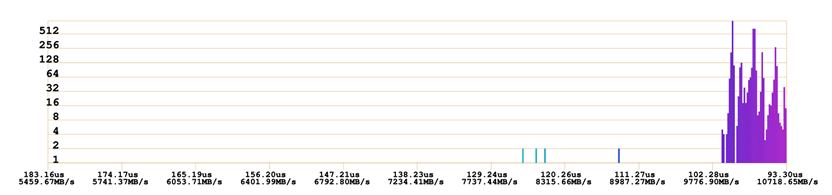}
    \caption{Histogram of the ping-pong times/speeds between all nodes using 1 MB packets in serial mode}
    \label{fig:MPI-Serial}
\end{figure}

\begin{figure}[h!]
    \centering
    \includegraphics[width=\columnwidth]{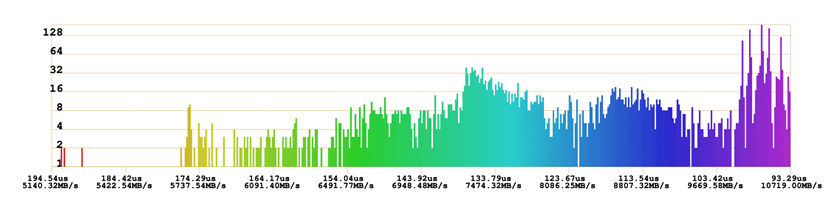}
    \caption{Histogram of the ping-pong times/speeds between all nodes using 1 MB packets in parallel mode}
    \label{fig:MPI-Parallel}
\end{figure}

The high performance cluster network has the Fat Tree topology and is build from six Mellanox SB7890 (unmanaged) and two SB7800 (managed) switches that provide 100 Gbit/s (IB EDR) connections between the nodes. The performance of the interconnect has been measured with the ``mpilink'' program that times the ping-pong exchange between each node~\cite{fzjmpilink}. To make the measurements we have installed Mellanox HPC package drivers and used openMPI version 3.1.2. The results are shown in fig.~\ref{fig:MPI-Serial} for serial mode runs and in fig.~\ref{fig:MPI-Parallel} for parallel mode runs. 


The serial mode sends packages to each node when previous communication has finished, while in parallel mode all sends and receives are issued at the same time. The parallel mode probes the package contention, while serial mode allows to establish the absolute speed and discover any failing links. The communication in serial mode is centered around the speed of 
$10.2 \pm 0.5 GB/s$. 
The parallel mode reveals certain over-subscription of the Fat Tree network~--- while the computational nodes are balanced the additional traffic from the file services causes delays in the transmission. This problem will be addressed in future upgrades.

\subsection{Operating System and cluster management} \label{sec:Luna}
The ``Zhores'' cluster is managed by~``Luna''~\cite{trinityX} provisioning tool which can be installed in~a~fault tolerant active-passive cluster setup with TrinityX platform. The Luna management system was developed by ClusterVision BV. The system automates the creation of all the services and cluster configuration that make a~bunch of~servers a~unified computational machine. 
\begin{figure}[h!]
    \centering
    \includegraphics[width=\columnwidth]{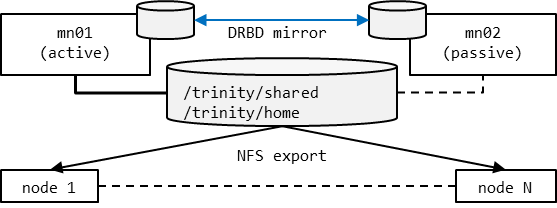}
    \caption{Organization of the ``Zhores'' cluster management with Luna system.}
    \label{fig:luna_failsafe}
\end{figure}

The~cluster management software supports the following essential features:
\begin{itemize}
    \item 
    All cluster configuration is kept in the Luna database and all cluster nodes boot from this information which is held in one place. This database is mirrored between the management nodes with the~DRBD filesystem and the active management node provides access to data for every node in the cluster with the~NFS share, see fig.~\ref{fig:luna_failsafe}. 
    \item
    Node provisioning from OS images is based on the BitTorrent protocol~\cite{bit-torrent} for efficient simultaneous (disk-less or standard) boot; the image management allows to grab an OS image from a running node to a file, clone images for testing or backup purposes; a group of nodes can use the same image for provisioning that fosters unification of cluster configuration. Nodes use PXE protocol to load a service image that implements the booting procedure.
    \item
    All the nodes (or groups of nodes) in the cluster can be switched on/off and reset with the IPMI protocol from the management nodes with a single command.
    \item
    The cluster services setup on the management node in a fault tolerant way include the following: DHCP, DNS, OpenLDAP, Slurm, Zabbix, Docker-repository, etc. 
\end{itemize}

The management nodes are based on CentOS~7.5 and force same OS on the compute nodes; additional packages, specific drivers and different kernel versions can be included in the images for the cluster nodes. The installation requires each node to have at least two ethernet network interfaces, one dedicated to the management traffic and the other used for administrative access.
A single cluster node can be booted within 2.5~minutes (over 1~GbE), and the whole "Zhores" cluster cold start takes 5~minutes to fully operational state.
 
\subsection{The queueing system} \label{sec:Queues}
Work queues have been organized with the Slurm workload manager to reflect the different application profiles of users of the cluster. Several nodes have been given to dedicated projects (gn26, anlab) and one CPU-only node is setup for debugging work (cn44). The remaining nodes have been combined in queues for the GPU-nodes (gn01--gn25) and for the CPU-nodes (cn01--cn43). 

\subsection{Linpack run}
The Linpack benchmark was performed as a part of the cluster evaluation procedure and to rate the supercomputer for the performance comparison. The results of the run are shown in table~\ref{tab:linpack} separately for the GPU and for all nodes using only CPU computation.
\begin{table}[h!]
    \centering
    \begin{tabular}{|c|>{\centering}p{12mm}|c|c|>{\centering}p{18mm}|>{\centering}p{16mm}|c|>{\centering}p{8mm}|}
    \hline
    Part       & \multirow{2}{*}{N} & NB  & T   & $R_{max}$ & $R_{peak}$ & eff. & P
    \tabularnewline
    nodes/core &   & P Q & [s] & [TFlop/s] & [TFlop/s] & [\%] & [kW]
    \tabularnewline
    \hline
    gn01-26 & \multirow{2}{*}{452352} & 192 & \multirow{2}{*}{124.4} & \multirow{2}{*}{$496(\pm2\%)$} & \multirow{2}{*}{811.2} & \multirow{2}{*}{61.1} & \multirow{2}{*}{48.9}
    \tabularnewline
    26/930     &   & 13 8 &    &           &            &       & 
    \tabularnewline
    \hline
    gn;cn;hn  & \multirow{2}{*}{\hspace{-1.5mm}1125888} & 384 & \multirow{2}{*}{7913.5} & \multirow{2}{*}{\hspace{-1.5mm}$120.2(\pm2\%)$} & \multirow{2}{*}{158.7} & \multirow{2}{*}{75.6} & \multirow{2}{*}{$35^*$} 
    \tabularnewline
    72/2028   &    & 12 12 &   &           &            &       & 
    \tabularnewline
    \hline
    \end{tabular}
    \caption{Linpack performance of the ``Zhores'' cluster run separately on the GPU nodes and with all CPU resources. The power draw for CPU Linpack run is estimated (*).}
    \label{tab:linpack}
\end{table}

``Zhores'' supercomputer is significant for the Russian computational science community and has reached position 6 in the Russian and CIS TOP-50 list~\cite{top50}.
\section{Applications}
\label{sec:apps}

\subsection{Algorithms for~aggregation and fragmentation equations}

In~our benchmarks we used parallel implementation of~efficient numerical methods for the aggregation and fragmentation equations \cite{matveev2015fast, krapivsky2010kinetic} and also parallel implementation of the solver for advection-driven coagulation process \cite{matveev2015parallel}. Its sequential version has already been utilized in a number of applications \cite{matveev2017oscillations, matveev2018anderson} and can be considered as one of the most efficient algorithms for a class of Smoluchowski-type aggregation kinetic equations. It is worth to stress that parallel algorithm for pure aggregation-fragmentation equations is based mostly on the performance of ClusterFFT operation which is a dominating operation in terms of algorithmic complexity, thus its scalability is extremely limited. Nevertheless for 128 cores we obtain speedup of calculations by more than 85 times, see table~\ref{tab:Parallel_smoluchowski}.

\begin{figure}[h!]
    \centering
    \includegraphics[width=\columnwidth]{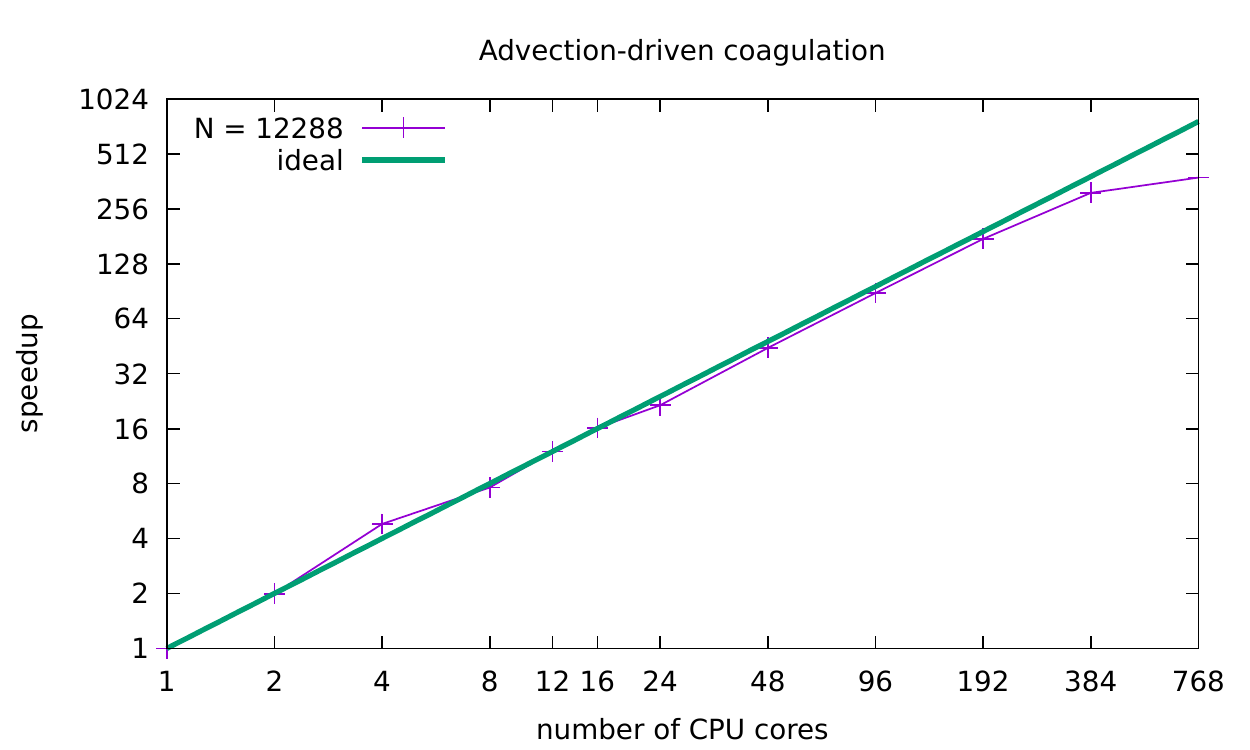}
    \caption{Parallel advection-driven aggregation solver on CPUs, Ballistic kernel, domain size~$N = 12288$.}
    \label{fig:adv_coag}
\end{figure}

In the case of the parallel solver, for advection-driven coagulation \cite{matveev2018parallel} we obtain almost ideal acceleration with utilization of the algorithm for almost full CPU-based segment. In this case, the algorithm is based on the one-dimensional domain decomposition along the spatial coordinate and has a very good scalability, see table~\ref{tab:my_label} and fig.~\ref{fig:adv_coag}. The experiments have been performed using Intel\textregistered{} compilers and the  Intel\textregistered{} MKL library.

\begin{table}[h!]
    \centering
    \begin{tabular}{cc}
    \hline
      Number of CPU cores & Time, sec
      \tabularnewline
      \hline
         1  & 585.90
         \tabularnewline
         2  & 291.69
         \tabularnewline
         4  & 152.60
         \tabularnewline
         8  & 75.60
         \tabularnewline
        16  & 41.51
        \tabularnewline
        32  & 20.34
        \tabularnewline
        64  & 12.02
        \tabularnewline
        128  & 6.84
        \tabularnewline
        \hline
    \end{tabular}
    \caption{Computational times for~16~time-integration steps for the parallel implementation of algorithm for the aggregation and fragmentation equations with~$N = 2^{22}$ strongly-coupled nonlinear~ODEs. In this benchmark we utilized the nodes from the CPU segment of the cluster.}
    \label{tab:Parallel_smoluchowski}
\end{table}


\begin{table}[h!]
    \centering
    \begin{tabular}{cc}
        \hline
        Number of CPU cores & Time, sec 
        \tabularnewline
        \hline
        1  &	1706.50
            \tabularnewline
        2  &	856.057
            \tabularnewline
        4  &	354.85
            \tabularnewline
        8  &	224.44
            \tabularnewline
        12 &	142.66
            \tabularnewline
        16 &	105.83
            \tabularnewline
        24 &	79.38
            \tabularnewline
        48 &	38.58
            \tabularnewline
        96 &	19.31
            \tabularnewline
        192 &	9.75
            \tabularnewline
        384 &	5.45
            \tabularnewline
        768 &	4.50
            \tabularnewline
        \hline
    \end{tabular}
    \caption{Parallel advection-coagulation solver on CPUs, Ballistic kernel, domain size $N  \times M = 12 288$, 16 time-integration steps. This benchmark utilized up to 32 nodes from the CPU segment of the cluster.}
    \label{tab:my_label}
\end{table}

\begin{figure}[h!]
    \centering
    \includegraphics[width=\columnwidth]{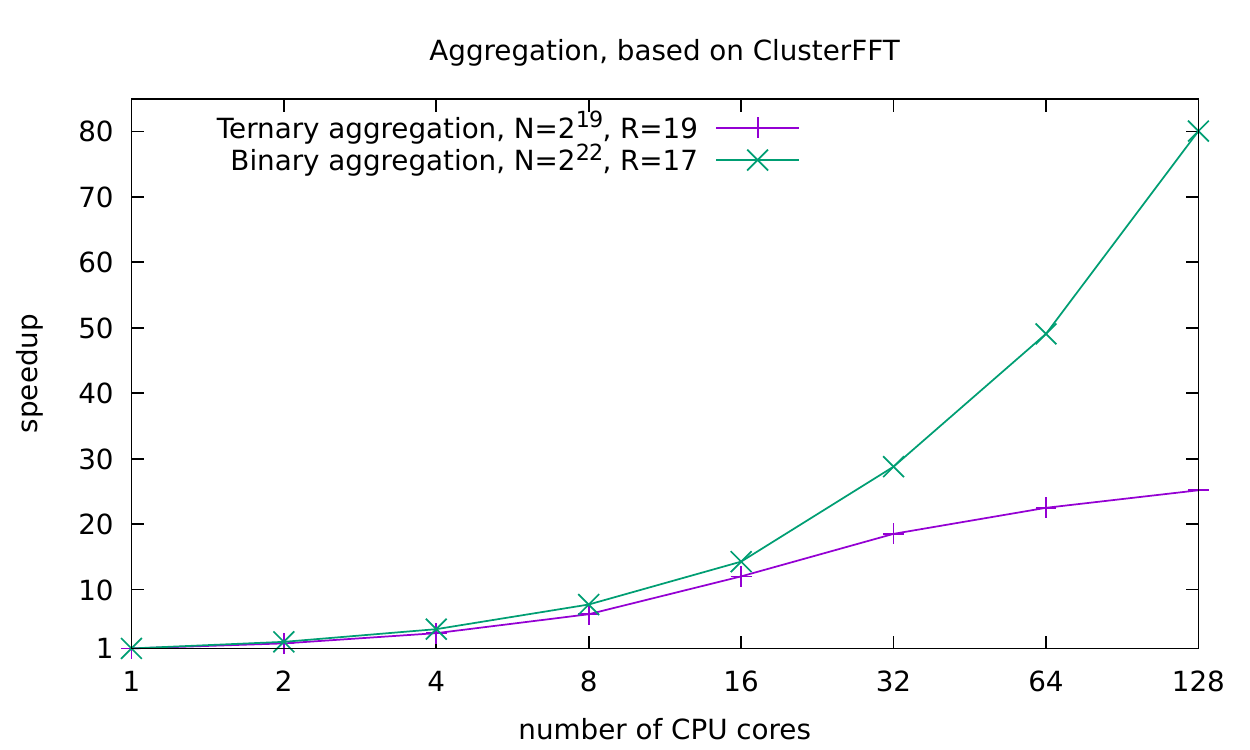}
    \caption{Parallel binary and ternary aggregation solvers on CPU, Ballistic-like kernels, 16 and 10 time-integration steps for~$N = 2^{22}$ and $N = 2^{19}$ nonlinear~ODEs, respectively. Parameter~$R$ denotes the~rank of used matrix and tensor decompositions.}
    \label{fig:aggr}
\end{figure}

Alongside with the consideration of the well-known two-particle problem of aggregation, we have measured the performance for~a~parallel implementation of~a~more~general three-particle~(ternary) Smoluchowski-type kinetic aggregation equations~\cite{stefonishin2018tensor}. In~this case the~algorithm is somewhat similar to~the~one for~standard binary aggregation. 
However the number of the floating point calculations and the size of the allocated memory increases as compared to the binary case, because the dimension of the low rank Tensor Train~(TT) decomposition~\cite{oseledets2010tt} is naturally bigger in ternary case.
The~most computationally expensive operation in~the parallel implementation of~the~algorithm is~also the ClusterFFT. The speedup of~the~parallel ternary aggregation algorithm applied to the empirically derived ballistic-like kinetic coefficients~\cite{matveev2018numerical} is shown in table~\ref{tab:ter_smoluchowski}. In~full accordance with~the~structure of~ClusterFFT and the problem complexity one needs to~increase the~parameter~$N$ of the used differential equations in~order~to~obtain scalability. Speedups for both implementations of binary and ternary aggregation are shown on fig.~\ref{fig:aggr}. The~experiments have been performed using Intel compilers and Intel MKL library.


\begin{table}[h!]
    \centering
    \begin{tabular}{cc}
    \hline
      Number of CPU cores & Time, sec
      \tabularnewline
      \hline
        1 & 624.19 
        \tabularnewline
        2 & 351.21
        \tabularnewline
        4 & 186.83
        \tabularnewline
        8 & 100.33
        \tabularnewline
        16 & 52.02
        \tabularnewline
        32 & 33.74
        \tabularnewline
        64 & 27.74
        \tabularnewline
        128 & 24.80
        \tabularnewline
        \hline
    \end{tabular}
    \caption{Computational times for~10~time-integration steps for~parallel implementation of the algorithm for~ternary aggregation equations with~$N = 2^{19}$ nonlinear~ODEs.}
    \label{tab:ter_smoluchowski}
\end{table}

\subsection{Gromacs}
Classical molecular dynamics is an effective method with high predictive ability in a wide range of scientific fields~\cite{kapral2005molecular,sutmann2002classical}. Using Gromacs~2018.3 software~\cite{abraham2015gromacs, van2005gromacs} we have performed molecular dynamics simulations in~order~to test the ``Zhores'' cluster performance. As~a~model system we chose 125~million Lennard-Jones spheres with the~Van der~Waals  cut-off radius of $1.2$~nm and with the Berendsen thermostat. All tests were conducted with a single precision version of Gromacs.

\begin{figure}[h!]
    \centering
    \includegraphics[width=\columnwidth]{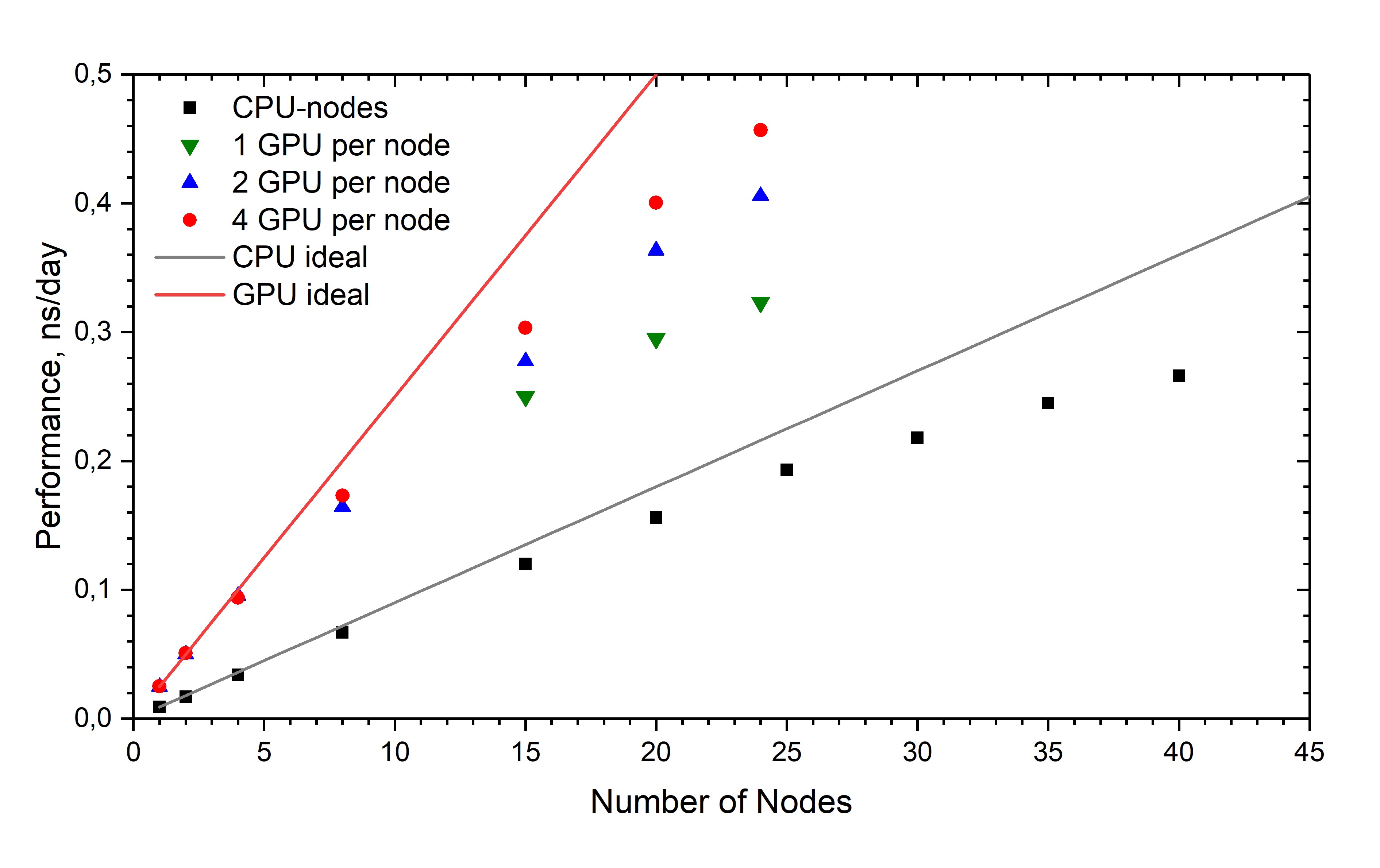}
    \caption{Performance of the molecular dynamic simulations of 125 million Lennard-Jones spheres using Gromacs 2018.3 as a function of nodes number. Note, that there are only 26 GPU nodes on the cluster.}
    \label{fig:gromacs_bench}
\end{figure}

The results are presented in fig.~\ref{fig:gromacs_bench}. We measured the performance as a function of the number of nodes; we have used up to~40 CPU nodes and up to~24 GPU nodes. We have used 4~OpenMP threads per MPI~process. Each task was performed 5 times with following averaging in order to obtain final performance.
Grey and red solid lines show linear acceleration of the program on CPU and GPU nodes, respectively. In case of the CPU-nodes, one can see almost ideal speedup. With a~large number of CPU-nodes, the speedup deviates from linear and grows slower.

\begin{figure}[h!]
    \centering
    \includegraphics[width=\columnwidth]{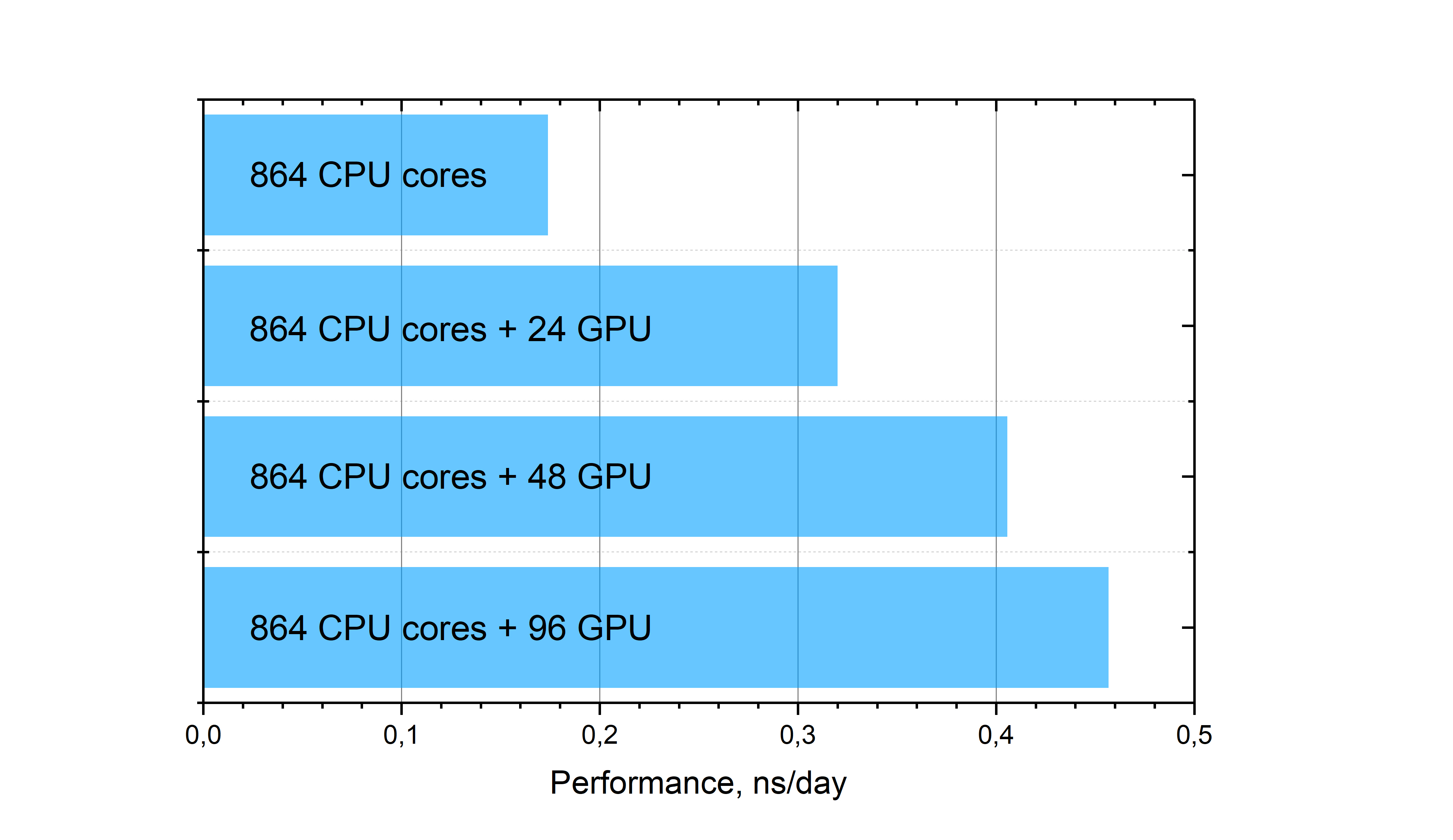}
    \caption{Performances for different configurations of 24 GPU-nodes: 0, 1, 2, and 4 GPU per node}
    \label{fig:gromacs_bench_bar}
\end{figure}

To~test performance on the GPU-nodes, we have performed simulations with 1, 2 and 4~graphics cards per~node. The use of all 4~graphical cards demonstrates good scalability, while 2~GPU per node shows slightly lower speedup. Runs with 1~GPU per node demonstrates worse performance, especially with high number of nodes. To compare the efficiency for different number GPU per node, we show the performance for the four configurations (0, 1, 2 and 4 GPU) using 24 GPU-nodes in fig.~\ref{fig:gromacs_bench_bar} as a~bar chart. The 4~GPU per node configuration gives about $2.5$~times higher performance than running the program only on the CPU cores. And even 1~GPU per node gives significant performance increase compared to the CPU only run.

\section{Neurohackathon at Skoltech}
\label{sec:Neurohackathon}
``Zhores" cluster was used as the main computing power during the ``Neurohackathon" in the field of~neuro-medicine, held in Skoltech from 16-th to 18-th of November 2018 under the umbrella of the National Technology Initiative. It consisted of~two~tracks: scientific and open. The~scientific track included the~solution of~the~tasks of~predictive analytics related to~the~analysis of~MRI~images of~the~brain of~patients containing changes characteristic of multiple sclerosis~(MS). This activity handled private data. Therefore special attention was paid to the IT~security. It was necessary to divide the cluster resources such that Skoltech continued its scientific activities while the hackathon participants competed transparently on this facility at the same time.

To address this problem, a two-stage authentication system was chosen using several levels of virtualization. Access to the cluster was made through the VPN tunnel using Cisco ASA and Cisco AnyConnect; then the SSH~(RFC 4251) protocol was used to access the consoles of the operating system~(OS) of~the~participants.

The~virtualization was provided at~the~level of~a~data network
through the~IEEE~802.1Q~(VLAN) protocol and OS level Docker~\cite{Docker} containerization with the ability to connect to~GPU accelerators. The container worked in its address space and
in~a~separate VLAN, so we achieved an additional isolation level from~the~host machine. Also at the Linux kernel level, the namespace feature was turned on and~the~user and group IDs were remapped to obfuscate the superuser rights on the host machine.

As a result, each participant of the Neurohackathon had
a~docker container with access via~the~SSH protocol to the console and used the~https protocol to Jupiter application on his VM. The four Nvidia Tesla V100 accelerators on the GPU nodes were used for the computing.

The number of teams participating in the competition had rapidly increased from 6 to 11 one hour before the start of the event. The usage of virtualization technology and the flexible architecture of the cluster allowed us to provide all teams with the necessary resources and start the hackathon on time. 

\section{Conclusions}
\label{sec:Conclusions}
In conclusion, we have presented the Petaflops supercomputer ``Zhores" installed in Skoltech CDISE that will be actively used for multidisciplinary research in data-driven simulations, machine learning, Big Data and artificial intelligence. Linpack benchmark placed this cluster at position 6 of the Russian and CIS TOP-50 Supercomputer list. Initial tests show a good scalability of the modeling applications and prove that the new computing instrument can be used to support advanced research at Skoltech and for all its research and industrial partners.

\section{Acknowledgements}
\label{Acknowledgements}
We are thankful to the Nobel Prize laureate Prof. Zhores Alferov for his agreement to lend his name to this project. Authors acknowledge valuable contribution from Dmitry Sivkov (Intel) and Sergey Kovylov (NVidia) for their help in running Linpack tests during the deployment of ``Zhores" cluster and Dmitry Nikitenko (MSU) for his help in filling the forms for the Russia and CIS Top-50 submission. We are indebted to Dr. Sergey Matveev for valuable consultations and Denis Shageev for indispensable help with the cluster installation. We would also like to thank Prof. Dmitry Dylov, Prof. Andrey Somov, Prof. Dmitry Lakontsev, Seraphim Novichkov and Eugene Bykov for their active role in organization of the Neurohackathon. We are also thankful to Prof. Ivan Oseledets and Prof. Eugene Burnaev and their team members for testing of the cluster.







\bibliographystyle{plain}
\bibliography{sample}

\begin{thebibliography}{10}

\bibitem{fzjmpilink}
Julich mpilinktest.
\newblock http://www.fz-juelich.de/jsc/linktest.
\newblock Accessed: 2018-12-15.

\bibitem{bit-torrent}
The {B}it{T}orrent {P}rotocol {S}pecification.
\newblock http://www.bittorrent.org, 2008.
\newblock Accessed: 2018-12-15.

\bibitem{COLFAX}
Capabilities of {I}ntel{\textregistered} {AVX}-512 in {I}ntel{\textregistered}
  {X}eon{\textregistered} {S}calable {P}rocessors (skylake).
\newblock https://colfaxresearch.com/skl-avx512, 2017.
\newblock Accessed: 2018-12-15.

\bibitem{Docker}
Docker.
\newblock http://www.docker.com, 2018.
\newblock Accessed: 2018-12-15.

\bibitem{Software_modules}
Environment {M}odules.
\newblock http://modules.sourceforge.net, 2018.
\newblock Accessed: 2018-12-15.

\bibitem{top50}
Top50 {S}upercomputers~(in~{R}ussian).
\newblock http://top50.supercomputers.ru, 2018.
\newblock Accessed: 2018-12-15.

\bibitem{abraham2015gromacs}
M.~J. Abraham, T.~Murtola, R.~Schulz, S.~P{\'a}ll, J.~C. Smith, B.~Hess, and
  E.~Lindahl.
\newblock Gromacs: High performance molecular simulations through multi-level
  parallelism from laptops to supercomputers.
\newblock {\em SoftwareX}, 1:19--25, 2015.

\bibitem{babenko2014neural}
A.~Babenko, A.~Slesarev, A.~Chigorin, and V.~Lempitsky.
\newblock Neural codes for image retrieval.
\newblock In {\em European conference on computer vision}, pages 584--599.
  Springer, 2014.

\bibitem{baez2018quantum}
J.~C. Baez and J.~D. Biamonte.
\newblock {\em Quantum Techniques in Stochastic Mechanics}.
\newblock World Scientific, 2018.

\bibitem{biamonte2017quantum}
J.~Biamonte, P.~Wittek, N.~Pancotti, P.~Rebentrost, N.~Wiebe, and S.~Lloyd.
\newblock Quantum machine learning.
\newblock {\em Nature}, 549(7671):195, 2017.

\bibitem{burkov2018deep}
E.~Burkov and V.~Lempitsky.
\newblock Deep neural networks with box convolutions.
\newblock In {\em Advances in Neural Information Processing Systems}, pages
  6212--6222, 2018.

\bibitem{trinityX}
ClusterVision BV.
\newblock Luna.
\newblock https://clustervision.com, 2017.
\newblock Accessed: 2018-12-15.

\bibitem{cao2018robust}
W.~Cao, K.~Wang, G.~Han, J.~Yao, and A.~Cichocki.
\newblock A robust {PCA} approach with noise structure learning and
  spatial-spectral low-rank modeling for hyperspectral image restoration.
\newblock {\em IEEE Journal of Selected Topics in Applied Earth Observations
  and Remote Sensing}, (99):1--17, 2018.

\bibitem{cichocki2018tensor}
A.~Cichocki.
\newblock Tensor networks for dimensionality reduction, big data and deep
  learning.
\newblock In {\em Advances in Data Analysis with Computational Intelligence
  Methods}, pages 3--49. Springer, 2018.

\bibitem{coles2017nanostructure}
S.~W. Coles, M.~Mishin, S.~Perkin, M.~V. Fedorov, and V.~B.
  Ivani{\v{s}}t{\v{s}}ev.
\newblock The nanostructure of a lithium glyme solvate ionic liquid at
  electrified interfaces.
\newblock {\em Physical Chemistry Chemical Physics}, 19(18):11004--11010, 2017.

\bibitem{docampo2016molecular}
B.~Docampo-{\'A}lvarez, V.~G{\'o}mez-Gonz{\'a}lez, H.~Montes-Campos, J.~M.
  Otero-Mato, T.~M{\'e}ndez-Morales, O.~Cabeza, L.~J. Gallego, R.~M.
  Lynden-Bell, V.~B. Ivani{\v{s}}t{\v{s}}ev, M.~V. Fedorov, et~al.
\newblock Molecular dynamics simulation of the behaviour of water in
  nano-confined ionic liquid-water mixtures.
\newblock {\em Journal of Physics: Condensed Matter}, 28(46):464001, 2016.

\bibitem{gomez2017molecular}
V.~G{\'o}mez-Gonz{\'a}lez, B.~Docampo-{\'A}lvarez, T.~M{\'e}ndez-Morales,
  O.~Cabeza, V.~B. Ivani{\v{s}}t{\v{s}}ev, M.~V. Fedorov, L.~J. Gallego, and
  L.~M. Varela.
\newblock Molecular dynamics simulation of the structure and interfacial free
  energy barriers of mixtures of ionic liquids and divalent salts near a
  graphene wall.
\newblock {\em Physical Chemistry Chemical Physics}, 19(1):846--853, 2017.

\bibitem{kapral2005molecular}
R.~Kapral and G.~Ciccotti.
\newblock Molecular dynamics: an account of its evolution.
\newblock In {\em Theory and Applications of Computational Chemistry}, pages
  425--441. Elsevier, 2005.

\bibitem{klyuchnikov2018data}
N.~Klyuchnikov, A.~Zaytsev, A.~Gruzdev, G.~Ovchinnikov, K.~Antipova,
  L.~Ismailova, E.~Muravleva, E.~Burnaev, A.~Semenikhin, A.~Cherepanov, et~al.
\newblock Data-driven model for the identification of the rock type at a
  drilling bit.
\newblock {\em arXiv preprint arXiv:1806.03218}, 2018.

\bibitem{kononenko2015learning}
D.~Kononenko and V.~Lempitsky.
\newblock Learning to look up: Realtime monocular gaze correction using machine
  learning.
\newblock In {\em Proceedings of the IEEE Conference on Computer Vision and
  Pattern Recognition}, pages 4667--4675, 2015.

\bibitem{kononenko2018semi}
D~Kononenko and V.~Lempitsky.
\newblock Semi-supervised learning for monocular gaze redirection.
\newblock In {\em Automatic Face \& Gesture Recognition (FG 2018), 2018 13th
  IEEE International Conference on}, pages 535--539. IEEE, 2018.

\bibitem{krapivsky2010kinetic}
P.~L. Krapivsky, S.~Redner, and E.~Ben-Naim.
\newblock {\em A kinetic view of statistical physics}.
\newblock Cambridge University Press, 2010.

\bibitem{LittleLaw}
J.~D. Little.
\newblock “a proof for the queueing formula: ${L}=\lambda \cdot {W}$".
\newblock {\em Operations Research}, 9(3), 1961.

\bibitem{matveev2015parallel}
S.~A. Matveev.
\newblock A parallel implementation of a fast method for solving the
  smoluchowski-type kinetic equations of aggregation and fragmentation
  processes.
\newblock {\em Vychislitel'nye Metody i~Programmirovanie (in~{R}ussian)},
  16(3):360--368, 2015.

\bibitem{matveev2017oscillations}
S.~A. Matveev, P.~L. Krapivsky, A.~P. Smirnov, E.~E. Tyrtyshnikov, and N.~V.
  Brilliantov.
\newblock Oscillations in aggregation-shattering processes.
\newblock {\em Physical review letters}, 119(26):260601, 2017.

\bibitem{matveev2015fast}
S.~A. Matveev, A.~P. Smirnov, and E.~E. Tyrtyshnikov.
\newblock A fast numerical method for the {C}auchy problem for the
  {S}moluchowski equation.
\newblock {\em Journal of Computational Physics}, 282:23--32, 2015.

\bibitem{matveev2018anderson}
S.~A. Matveev, V.~I. Stadnichuk, E.~E. Tyrtyshnikov, A.~P. Smirnov, N.~V.
  Ampilogova, and N.~V. Brilliantov.
\newblock Anderson acceleration method of finding steady-state particle size
  distribution for a wide class of aggregation--fragmentation models.
\newblock {\em Computer Physics Communications}, 224:154--163, 2018.

\bibitem{matveev2018numerical}
S.~A. Matveev, D.~A. Stefonishin, A.~P. Smirnov, A.~A. Sorokin, and E.~E.
  Tyrtyshnikov.
\newblock Numerical studies of solutions for kinetic equations with
  many-particle collisions.
\newblock In {\em Journal of Physics: Conference Series}. IOP Publishing,
  accepted, in press.

\bibitem{matveev2018parallel}
S.~A. Matveev, R.~R. Zagidullin, A.~P. Smirnov, and E.~E. Tyrtyshnikov.
\newblock Parallel numerical algorithm for solving advection equation for
  coagulating particles.
\newblock {\em Supercomputing Frontiers and Innovations}, 5(2):43--54, 2018.

\bibitem{STREAMS}
J.~D. McCalpin.
\newblock Memory bandwidth and machine balance in current high performance
  computers.
\newblock {\em IEEE Computer Society Technical Committee on Computer
  Architecture (TCCA) Newsletter}, pages 19--25, dec 1995.

\bibitem{LMBench}
L.~W. McVoy and C.~Staelin.
\newblock lmbench: Portable tools for performance analysis.
\newblock In {\em Proceedings of the {USENIX} Annual Technical Conference, San
  Diego, California, USA, January 22-26, 1996}, pages 279--294, 1996.

\bibitem{mirvakhabova2018field}
L.~Mirvakhabova, M.~Pukalchik, S.~Matveev, P.~Tregubova, and I.~Oseledets.
\newblock Field heterogeneity detection based on the modified fastica
  {RGB}-image processing.
\newblock In {\em Journal of Physics: Conference Series}, volume 1117, page
  012009. IOP Publishing, 2018.

\bibitem{muravleva2018application}
E.~Muravleva, I.~Oseledets, and D.~Koroteev.
\newblock Application of machine learning to viscoplastic flow modeling.
\newblock {\em Physics of Fluids}, 30(10):103102, 2018.

\bibitem{novikov2018satellite}
G.~Novikov, A.~Trekin, G.~Potapov, V.~Ignatiev, and E.~Burnaev.
\newblock Satellite imagery analysis for operational damage assessment in
  emergency situations.
\newblock In {\em International Conference on Business Information Systems},
  pages 347--358. Springer, 2018.

\bibitem{novikov2018improving}
I.~S. Novikov and A.~V. Shapeev.
\newblock Improving accuracy of interatomic potentials: more physics or more
  data? {A} case study of silica.
\newblock {\em Materials Today Communications}, 2018.

\bibitem{oseledets2010tt}
I.~Oseledets and E.~Tyrtyshnikov.
\newblock T{T}-cross approximation for~multidimensional arrays.
\newblock {\em Linear Algebra and its Applications}, 432(1):70--88, 2010.

\bibitem{somov2018bacteria}
A.~Somov, P.~Gotovtsev, A.~Dyakov, A.~Alenicheva, Y.~Plehanova, S.~Tarasov, and
  A.~Reshetilov.
\newblock Bacteria to power the smart sensor applications: Biofuel cell for
  low-power iot devices.
\newblock In {\em 4th World Forum on Internet of Things (WF-IoT)}, pages
  802--806. IEEE, 2018.

\bibitem{somov2018pervasive}
A.~Somov, D.~Shadrin, I.~Fastovets, A.~Nikitin, S.~Matveev, I.~Oseledets, and
  O.~Hrinchuk.
\newblock Pervasive agriculture: {I}o{T} enabled greenhouse for plant growth
  control.
\newblock {\em IEEE Pervasive Computing Magazine}, 2018.

\bibitem{0953-8984-30-32-32LT03}
Sergey Sosnin, Maksim Misin, David~S Palmer, and Maxim~V Fedorov.
\newblock 3d matters! 3d-rism and 3d convolutional neural network for accurate
  bioaccumulation prediction.
\newblock {\em Journal of Physics: Condensed Matter}, 30(32):32LT03, 2018.

\bibitem{stefonishin2018tensor}
D.~A. Stefonishin, S.~A. Matveev, A.~P. Smirnov, and E.~E. Tyrtyshnikov.
\newblock Tensor decompositions for solving the equations of mathematical
  models of aggregation with multiple collisions of particles.
\newblock {\em Vychislitel'nye Metody i~Programmirovanie (in~{R}ussian)},
  19(4):390--404, 2018.

\bibitem{sutmann2002classical}
G.~Sutmann.
\newblock {\em Classical molecular dynamics and parallel computing}.
\newblock FZJ-ZAM, 2002.

\bibitem{ulyanov2018deep}
D.~Ulyanov, A.~Vedaldi, and V.~Lempitsky.
\newblock Deep image prior.
\newblock In {\em IEEE Conf. Computer Vision and Pattern Recognition (CVPR)},
  pages 9446--9454, 2018.

\bibitem{van2005gromacs}
D.~Van Der~Spoel, E.~Lindahl, B.~Hess, G.~Groenhof, A.~E. Mark, and H.~J.~C.
  Berendsen.
\newblock Gromacs: fast, flexible, and free.
\newblock {\em Journal of computational chemistry}, 26(16):1701--1718, 2005.

\bibitem{yokota2018semi}
T.~Yokota, Z.~R. Struzik, P.~Jurica, M.~Horiuchi, S.~Hiroyama, J.~Li,
  Y.~Takahara, K.~Ogawa, K.~Nishitomi, M.~Hasegawa, et~al.
\newblock Semi-automated biomarker discovery from pharmacodynamic effects on
  {EEG} in {ADHD} rodent models.
\newblock {\em Scientific reports}, 8(1):5202, 2018.

\bibitem{yurchenko2017parsing}
V.~Yurchenko and V.~Lempitsky.
\newblock Parsing images of overlapping organisms with deep singling-out
  networks.
\newblock In {\em CVPR}, pages 4752--4760, 2017.

\bibitem{zhang2018multi}
Y.~Zhang, Y.~Wang, G.~Zhou, J.~Jin, B.~Wang, X.~Wang, and A.~Cichocki.
\newblock Multi-kernel extreme learning machine for {EEG} classification in
  brain-computer interfaces.
\newblock {\em Expert Systems with Applications}, 96:302--310, 2018.

\end{thebibliography}





\end{document}